\documentclass[aps,prl,superscriptaddress,showpacs,twocolumn]{revtex4}
\usepackage{graphicx}
\begin{document}
\title{Ground state order and spin-lattice coupling in tetrahedral 
spin systems Cu$_2$Te$_2$O$_5$X$_2$}
\author{M.Prester}
  \email{prester@ifs.hr}
\affiliation{Institute of Physics, P.O.B.304, HR-10 000, Zagreb,
Croatia}

\author{A.Smontara}
  \email{ana@ifs.hr}
\affiliation{Institute of Physics, P.O.B.304, HR-10 000, Zagreb,
Croatia}

\author{I.\v Zivkovi\'c}
\affiliation{Institute of Physics, P.O.B.304, HR-10 000, Zagreb,
Croatia}

\author{A.Bilu\v si\'c}
\affiliation{Institute of Physics, P.O.B.304, HR-10 000, Zagreb,
Croatia}

\author{D.Drobac}
\affiliation{Institute of Physics, P.O.B.304, HR-10 000, Zagreb,
Croatia}

\author{H.Berger}
\affiliation{Institut de Physique de la Mati\`ere Complexe, EPFL,
CH-1015 Lausanne, Switzerland}

\author{F.Bussy}
\affiliation{Institute of Mineralogy and Geochemistry BFSH-2,
University of Lausanne, CH-1015 Lausanne, Switzerland}

\date{\today}

\begin{abstract}
High-resolution ac susceptibility and thermal conductivity measurement on 
Cu$_2$Te$_2$O$_5$X$_2$ (X=Br,Cl) single crystals are reported. For Br-sample, sample dependence
prevents to distinguish between possibilities of magnetically ordered and spin-singlet
ground states. In Cl-sample a three-dimensional transition at 18.5 K is accompanied by 
almost isotropic behavior of susceptibility and almost switching behavior of thermal conductivity.
Thermal conductivity studies suggest the presence of a tremendous spin-lattice 
coupling characterizing Cl- but not Br-sample. Below the transition
Cl-sample is in a complex magnetic state involving AF order but also the elements 
consistent with the presence of a gap in the excitation spectrum.
\end{abstract}

\pacs{775.10.Jm, 75.40.Cx, 75.45.+j}

\maketitle

In quantum magnets a nonmagnetic spin singlet ground state, an intriguing hallmark 
of their quantum nature, and a 
sizeable reduction of the range of magnetic correlations,
are two inseparable  phenomena \cite{mil1}. Thus, unlike classical spin systems, 
characterized by some kind of low-temperature magnetic long range order,
quantum magnetic systems may reveal just a short range ordered -a spin liquid- 
ground state \cite{mil1}, separated by a spin gap from the excitation spectrum.
However, the presence of a gap 
in properties of real physical quantum system,  does not necessarily 
grant a short-range ordered nonmagnetic singlet
ground state: from various reasons a long-range ordered 
ground states are even more frequent in known quantum magnets \cite{lem1}. 

Addressing the general problem of ground-state order we report in this work our 
studies on recently discovered \cite{joh} 
copper telluride Cu$_2$Te$_2$O$_5$X$_2$ systems 
(where X=Cl or Br), attracting a lot of attention \cite{lem2,gro,bre}. These systems 
belong to a category of Cu$^{2+}$, S=1/2, quantum magnets featuring antiferromagnetic 
(AF) Heisenberg interaction and a rich variety of quantum spin phenomena \cite{lem1}. 
Both Cu$_2$Te$_2$O$_5$Br$_2$ and Cu$_2$Te$_2$O$_5$Cl$_2$ integrates a common magnetic 
building block, tetrahedral clusters of S=1/2 spins. 
The Cu$_2$Te$_2$O$_5$X$_2$ systems thus combine intrinsic magnetic low-dimensionality (related to 
the quasi-zero-dimensional contribution of
weakly interacting clusters) and built-in frustration (related to tetrahedral topology). 
There is a common opinion that 
the full spectrum of properties may be reconstructed by 
including additional -inter-tetrahedral- magnetic interactions into 
the isolated-clusters model Hamiltonian \cite{gro,bre}. 
Even unperturbed, the model of
isolated clusters comprises intriguing 
excitation spectrum \cite{joh,lem2}: 
Its ground state is always a singlet (involving a quadrumer of all four spins 
or a product of the two 
individual dimers) while the first excited state can be either singlet or triplet, 
depending on the relative sizes of the two involved exchange interactions $J_1,J_2$. 
Experimentally supported \cite{joh} equality of these interactions, $J_1=J_2=J$, 
leads to a double degenerated singlet ground state, 
separated from the lowest excitation by a gap $\Delta=J$. 

Although Cu$_2$Te$_2$O$_5$Cl$_2$ and Cu$_2$Te$_2$O$_5$Br$_2$ samples are 
isostructural their low-temperature properties 
are significantly different \cite{lem2}. The most pronounced reported 
difference \cite{lem2} is that the Cl-system develops a 3D magnetic order 
below about 18 K while the Br-system builds up, below 11 K, a phase revealing 
specific, Raman active low-lying 
longitudinal magnetic modes \cite{lem2}. However, the specific nature and details
of both of the mentioned transitions/transformations have not been clarified enough
by previous studies \cite{joh,lem2,gro}.

Targeting the problem of intrinsic ground state and its order, in this 
work we report ac susceptibility and thermal conductivity studies of high-quality 
Cu$_2$Te$_2$O$_5$X$_2$ single crystals. AC susceptibility was measured using a commercial
CryoBIND set-up. The apparatus reaches its high resolution (better than $2 \cdot 10^{-9}$ emu)
employing measuring fields of the order of 1 Oe only. Low measuring fields are advantageous 
in studies of {\em spontaneous} magnetic ordering. The field level of a few Oersteds is 
three to four orders of magnitudes smaller than the typical field values 
used in previous DC-SQUID studies on powder \cite{joh,lem2} and
single-crystalline \cite{gro} forms of Cu$_2$Te$_2$O$_5$X$_2$ samples. 
Under conditions of our measurements we are thus pretty confident that the 
low-temperature behavior we report on in this work originates from the evolution of intrinsic  
ground state of Cu$_2$Te$_2$O$_5$X$_2$.

The single crystals we used were grown by the usual halogen vapor transport 
technique, using TeCl$_4$, Cl$_2$ or TeBr$_4$ as transport agents.
Semi-transparent dark- (Cu$_2$Te$_2$O$_5$Br$_2$) or light-green 
(Cu$_2$Te$_2$O$_5$Cl$_2$) samples
grow as needle-like single crystals with the apparent chain morphology.
The stoichiometry of the obtained single crystals were quantitatively probed by 
electron-probe micro analysis. This analysis
identified a good stoichiometry in all constituents (including Cl and Br), allowing only 
for a probability of small (1-2\%) systematic
under-stoichiometry of copper.

\begin{figure}[b]
\includegraphics[width=7.8cm,clip]{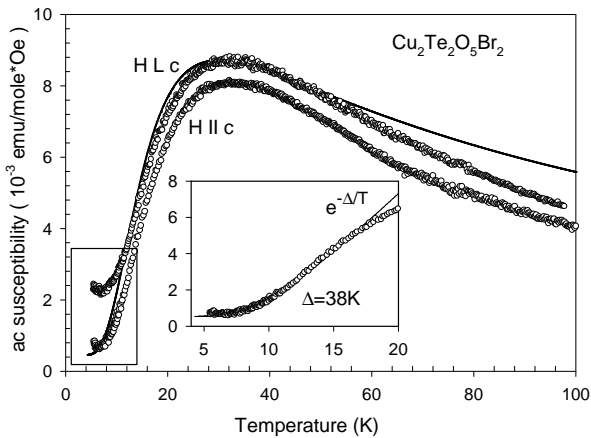}
\caption{AC susceptibility of one Cu$_2$Te$_2$O$_5$Br$_2$ 
single crystal (m=6.7 mg). Measurements were taken in two orientations 
to applied measuring field H$_{ac}$=2 Oe, at frequency 433 Hz.
Rectangular area designates the range of a sample-dependent saturation levels 
reached in measurements on different samples. 
Thin solid line plots the isolated-tetrahedra model from Ref. 3, taken with
the choice $J_1=J_2=48 K$. Inset demonstrates full compatibility with 
the elementary spin-gap expression (thin solid line).}
\end{figure}

The results of our ac susceptibility studies on Br- and Cl-single crystals are shown 
in Fig.1 and Fig.2, respectively. The positions of the susceptibility 
maxima coincide with the values 
reported previously \cite{joh,lem2}. Illustrating the quality of the single crystals 
used in our measurements, we point out that 
no sizeable Curie-like upturn (related to impurities and/or unpaired spins) has been observed 
down to 1.5 K. Our results show that below the respective 
maxima the susceptibility behavior 
of Br- and Cl-samples are very different. 
Measurements on Cu$_2$Te$_2$O$_5$Br$_2$ samples (Fig.1) 
reveals exponentially decreasing featureless susceptibility.
\begin{figure}[b]
\includegraphics[width=7.8cm,clip]{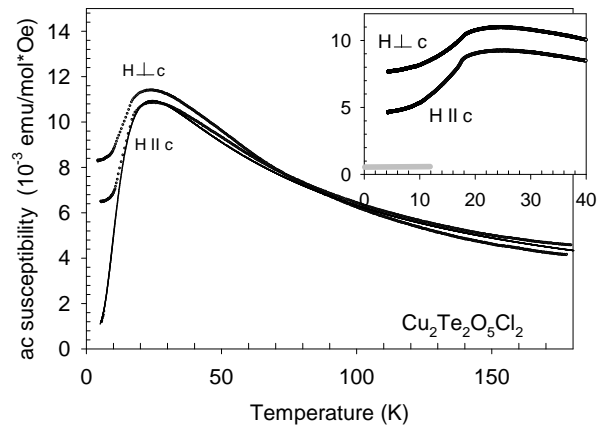}
\caption{AC susceptibility of one Cu$_2$Te$_2$O$_5$Cl$_2$ single crystal (m=17.6 mg). 
Measurements were taken in three orientations to applied measuring field H$_{ac}$=2 Oe, 
at frequency of 433 Hz. The result for the third orientation was very similar and 
has been omitted for clarity.
Thin solid line plots the isolated-tetrahedra model from Ref.3, taken with
the same choice $J_1=J_2=38.5 K$. Magnetic transition region is shown in the Inset. 
Thick gray line marks the position of the temperature-independent susceptibility part, as
determined in Ref.3.}
\end{figure}
At low temperatures susceptibility saturates at different levels depending, 
in most of the measured samples, on sample-to-applied field orientation.
While in `parallel' orientation (c-axis collinear with magnetic field direction)
the saturation level was found close to the value of orbital susceptibility 
(sum of the Van Vleck paramagnetism and core electron diamagnetism) in `orthogonal'
geometry saturation exhibits pronounced sample dependence.
Apart from evolution of susceptibility anisotropy, in our 
low-field measurements on Cu$_2$Te$_2$O$_5$Br$_2$
no explicit sign of magnetic ordering, in the form of a kink
or any other distinct susceptibility feature, could be 
detected in the whole temperature range below the susceptibility maximum. 

In contrast to Br-sample, in Cu$_2$Te$_2$O$_5$Cl$_2$ there is a sharp, almost isotropic, 
kink at T$_c$=18.5 K followed by an 
exponential susceptibility decay down to the relatively high level of
low temperature susceptibility saturation, Fig.2. 
However, we note that fine details of our results, like the mentioned 
level of low-temperature saturation, are somewhat sample- and thermal- (and/or magnetic) 
history-dependent. The latter dependence could probably 
be related to extrinsic magnetic contributions \cite{fus3}.

Verifying compatibility of our results 
with those published earlier we found out that the suggested 
susceptibility form for isolated tetrahedra (Figs.1,2, thin line) of Ref.3 
describes our data reasonably well. In fact, there is a perfect fit to the results for 
Cl-samples (Fig.2), using the same choice of interaction parameters as those identified 
earlier \cite{joh}, i.e., $J_1=J_2$=38.5 K. Quantitative accordance with the results for Br-sample
is not that good. Naturally, one can interpret the quantitative deviation from the
model prediction as an evidence of the inter-terahedral corrections to 
the unperturbed model Hamiltonian.

Thermal conductivity was measured along the long sample axis (c-axis). 
Magnetic susceptibility and thermal conductivity were measured on 
the samples from the same batch.
The results for Cu$_2$Te$_2$O$_5$Br$_2$ 
and Cu$_2$Te$_2$O$_5$Cl$_2$ samples are shown in Fig.3, 
covering the temperature range 8 K-150 K. (8 K represents
the lower margin of the temperature range of our set-up). Thermal conductivity reveals 
even more striking differences
between the two compounds. By lowering temperature 
thermal conductivity of Cu$_2$Te$_2$O$_5$Br$_2$ just monotonously increases 
forming a characteristic 
low-temperature maximum, typical for phonon thermal transport in
crystalline solids. In contrast, thermal conductivity of Cu$_2$Te$_2$O$_5$Cl$_2$, 
showing up a similar value 
and temperature dependence above 150 K as the Br-sample, first 
anomalously levels-off and saturates for temperatures below 40 K and then, 
below 15 K, very sharply increases and
approaches the respective thermal conductivity value of the Br-sample.

We first discuss our results for Cu$_2$Te$_2$O$_5$Br$_2$. 
Vanishing susceptibility shown in Fig.1 for parallel geometry
usually demonstrates a spin singlet ground state stabilizing in a compound: 
according to best of our knowledge it would be the first known spin singlet among 
the tetrahedral S=1/2 systems. In this case
the singlet state would rely on tetrahedral quadrumers (or dimer products, depending on 
$J_1$ vs. $J_2$ relationship). Limiting the temperature 
interval arbitrarily to the temperature range (4.2 K-16 K) 
a fit to the generic gap form $e^{-\Delta /T}$ identifies the spin gap value of $\Delta=38K$.
However, in orthogonal geometry 
susceptibility saturates at elevated but sample-dependent levels approaching
the vanishing level characterizing parallel-geometry in cases of one or two measured samples only. 
Elevated susceptibility
in any geometry is of course inconsistent with spin-singlet ground state.
The susceptibility anisotropy shown in Fig.1 is, on the other hand, fully compatible 
with magnetic ordering scheme proposed by Jensen et al. \cite{gro} thus the result for this 
particular sample would favor magnetically ordered ground state.
Still, due to mentioned sample dependence the question of ground state 
of Br-compound is not entirely resolved as yet. 
In our opinion a small amount of nonmagnetic impurities, possibly present in 
our samples \cite{fus3}, might play a crucial
role in stabilization of a particular (magnetic or non-magnetic)
ground state at low temperatures.

Focusing Cu$_2$Te$_2$O$_5$Cl$_2$ we discuss now its intriguing susceptibility and thermal
conductivity behavior below 18 K. The susceptibility kink was observed 
previously \cite{lem2} and our low field ac susceptibility measurements on single crystals 
confirm that there is indeed a spontaneous 3D transition underlying the kink feature. 
In vast majority of quantum magnets 
3d transition stabilizes the antiferromagnetically ordered ground state: only in CuGeO$_3$  
spin-Peierls mechanism \cite{has}, that involves also a symmetry breaking, 
stabilizes a dimerized, non-magnetic singlet state. The respective magnetic excitation 
spectrum can be either gapless or gapped,
in cases of magnetically ordered or spin singlet-ground states, respectively. 
(Of course, a gap can be introduced into the magnetic spectrum of magnetically 
ordered systems by presence other effects, like
magnetic anysotropy, into the effective Hamiltinian). However, 
as shown by numerous studies \cite{has} on doped CuGeO$_3$ these two
ground states are not entirely exclusive and locally coexisting AF ordered and dimerized phases 
cannot be excluded as well. Positioning the nature of the transition in the Cl-compund 
inside these possibilities \cite{fus2} is not an easy task. 
Discussing first the possibility of a long range AF order one 
immediately notes (Fig.2) that our susceptibility results cast some
doubts at least about the classical uniaxial N\'eel transition scenario: 
there is almost isotropic susceptibility drop 
below  T$_c$=18.5 K. (Generally, in an AF transition 
there is a sizeable susceptibility decrease along one (easy) magnetic axis 
while in the two orthogonal orientations susceptibility hardly changes). Indeed, an 
independent susceptibility anisotropy study by torque magnetometry \cite{milj} identified
the presence of magnetic ordering below 18.5 K; the order is however substantially 
more complex than the standard uniaxial N\'eel one. Also, a spin
flop phenomenon, a decisive feature of AF-ordered substances, 
could not be identified in these studies. Noteworthy, Cl-compound was found magnetically 
inhomogeneous \cite{fus3} thus making it difficult 
at present to separate intrinsic from extrinsic components in low temperature magnetism.

As far as the possibility of spin-Peierls transition is concerned
one first notes that the observed isotropy of ac susceptibility would be 
consistent with this type of transition:
isotropic susceptiility has been observed \cite{has} in spin-Peierls transition of CuGeO$_3$ and 
spin-Peierls-like transition \cite{iso} of NaV$_2$O$_5$. 
From Fig.2 one however realizes that the value of temperature independent 
orbital susceptibility is obviously much smaller than the low temperature susceptibility 
saturation in either directions, inconsistent with nonmagnetic spin-singlet ground state.

On the other hand the thermal conductivity anomaly (Fig.3) would be difficult to interpret 
without the participation of a gap in magnetic excitation spectrum of Cl-compound. 
Namely, there are generally 
very little changes of thermal conductivity at
(or in vicinity of) N\'eel transition point of AF substances. There are however several 
examples of pronounced anomalies either due scattering on critical fluctuations (e.g., in CoF$_2$) 
or magnon-phonon interaction (e.g., in FeCl$_2$). Still, to the best of our knowledge 
there are no examples of AF transition underlying the switching behavior of 
thermal conductivity like the one we report on for Cu$_2$Te$_2$O$_5$Cl$_2$. 
\begin{figure}[t]
\includegraphics[width=7.2cm,clip]{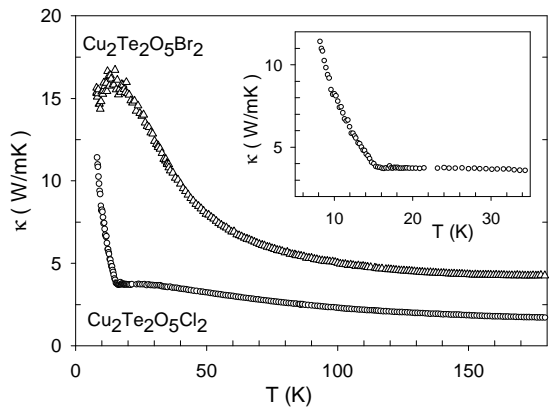}
\caption{Thermal conductivity of Cu$_2$Te$_2$O$_5$Cl$_2$ and Cu$_2$Te$_2$O$_5$Br$_2$ 
single crystals. A sharp increase of thermal conductivity of the Cu$_2$Te$_2$O$_5$Cl$_2$ 
sample, shown in the Inset on expanded scale, takes place below the 3D ordering 
temperature (see text).}
\end{figure}
Instead, the assumption of a gap enables more natural interpretation of our
thermal conductivity results (Fig.3). Br-sample reveals a typical phononic 
thermal transport down to low temperatures. Pronounced sample dependence, indicating
the presence of 3D transition below 11 K in some samples, has been also observed. 
No evidence for a combined (phonon+magnon) 
thermal transport or pronounced phonon scattering on magnons can be identified 
in our result \cite{fus1}. 
One concludes therefore that there is intrinsically small magneto-elastic coupling, 
as well as spin-phonon scattering, characterizing the Br-compound. Surprisingly,
the opposite is true for Cu$_2$Te$_2$O$_5$Cl$_2$: Fig.3 shows that the spin-lattice coupling is
unexpectedly large in this compound. (Unexpectedly, because of isomorphic structure of the
two compounds). In interpreting the result for Cu$_2$Te$_2$O$_5$Cl$_2$ one could naively assume 
that the abrupt and sizeable (factor of 4) thermal conductivity enhancement below 15 K reflects 
opening of an additional -magnetic- channel for thermal transport. However, a simple correlation
with the corresponding result for Cu$_2$Te$_2$O$_5$Br$_2$, which represents a natural `unperturbed' 
reference system, tells that the latter interpretation is very unlikely: there is 
obviously a mechanism that drastically suppress phonon conduction in intermediate 
temperature range but being very efficiently swept away below susceptibility kink -3D ordering- 
temperature T$_c$. 
We find a pronounced spin-phonon scattering responsible for the latter mechanism. The 
scattering is effective as long as there are magnetic excitations. Thus the assumption of a gap 
in the excitation spectrum of Cl-compound provide a simple interpretation for the explosive 
growth of thermal conductivity below the mid-gap temperatures.
Very similar interpretation and almost identical experimental observation has been reported
for spin-Peierls-like ordering \cite{vas2} in NaV$_2$O$_5$, as well for thermal 
transport \cite{hof,vas1} in ground-state-singlet system SrCu$_2$(BO$_3$)$_2$. 
Compared to NaV$_2$O$_5$ there are however significant differences: charge 
ordering (of V$^{5+}$ and V$^{4+}$) plays the main role in 
low temperature dynamics of the latter system.
In Cu$_2$Te$_2$O$_5$Cl$_2$ we find no grounds for charge ordering of this sort. 
We suggest therefore that a 
strong spin-lattice coupling represents an important ingredient in the 3D 
transition of this compound.
Microscopic background and details of the transition are unknown as yet.

In summary, pronounced sample dependence of Cu$_2$Te$_2$O$_5$Br$_2$ samples
precludes to resolve clearly between the possibilities of magnetically
ordered and spin-singlet ground states. An intriguing 3D transition, involving 
equally the spins and their couplings to lattice, stabilizes a complex 
low temperature magnetic state 
of Cu$_2$Te$_2$O$_5$Cl$_2$. As there are evidences for both, long range AF 
magnetic order and the presence
of a gap, our results suggest coexisting magnetically ordered and disordered 
spin-singlet domains characterizing
Cu$_2$Te$_2$O$_5$Cl$_2$ at low temperatures.

M.P. is grateful to M.Miljak, F.Mila and P.Lemmens for discussions. 
Support of the SCOPES program of 
the Swiss NSF is gratefully acknowledged. The sample preparation in Lausanne was 
supported by the NCCR research pool MaNEP of the Swiss NSF.


\begin{thebibliography}{99}
\bibitem{fus3}
Torque magnetometry studies \cite{milj} indicate the presence of 
magnetically ordered superparamagnetic inclusions in the samples
and their independent low temperature contributions to susceptibility.
Most probably, these inclusions originates from
small under-stoichiometry of copper in our samples, as verified by EPMA.
\bibitem{fus2}
A simple structural transition is very improbable due to 
smooth susceptibility variation and small but undoubtedly present
T$_c$ -dependence on the intensity of applied dc field \cite{lem2}.
\bibitem{fus1}
There is a reach history of evidences that both possible roles of magnons 
can be equally effective in thermal transport in various 
low dimensional spin gap systems.
For a recent reference, see Ref.\onlinecite{sol}.
\end{thebibliography}
\end{document}